\documentclass[,final]{aipproc}
\usepackage{graphics, epsfig}
\layoutstyle{6x9}

\begin{document}

\title{Binaries, cluster dynamics and population studies of stars and stellar phenomena}

\author{Dany Vanbeveren}{
  address={Astrophysical Institute, Vrije Universiteit Brussel, Pleinlaan 2, 1050 Brussels, Belgium.}
}

\begin{abstract}
 \noindent The effects of binaries on population studies of stars and stellar phenomena have been investigated over the 
past 3 decades by many research groups. Here we will focus mainly on the work that has been done recently in Brussels
and we will consider the following topics: the effect of binaries on overall galactic chemical evolutionary models and
on the rates of different types of supernova, the population of point-like X-ray sources where we distinguish the
standard high mass X-ray binaries and the ULXs, a UFO-scenario for the formation of WR+OB binaries in dense star
systems. Finally we critically discuss the possible effect of rotation on population studies.

\end{abstract}

\maketitle

\section{1. Introduction}

A population synthesis code calculates the temporal evolution of a population of stars in regions where star formation
is continuous in time or in starbursts. Population number synthesis (PNS) predicts the number of stars of a certain type
whereas population spectral synthesis (PSS) computes the effects that they have on the integrated spectrum. 

Close binary evolution has been a main research topic in Brussels for about 30 years. The last decade, we
investigated the effects of binaries on various aspects of population studies and we recently started to implement
stellar dynamics (using direct N-body integration techniques) in our PNS and PSS codes. In the present paper we will
summarize some of our recent results.   

\section{2. The effects of binaries on galactic evolution} 

{\bf Galactic chemical evolution (GCE).} Although most of the existing GCE codes account in some parametrized
way for the effects of supernova (SN) explosions of type Ia (SN Ia), they do not account in a consistent way for the
evolution of the whole population of binaries. I sometimes get the impression that it is not always realized that the
observed SN Ia rate indicates indirectly that a significant fraction of the intermediate mass stars are interacting
binary members, or that the observations of OB populations including X-ray binaries, double neutron star binaries or SN
type Ib indicate that also the massive close binary frequency is large.    

In order to study the effects of binaries on GCE, one has to combine PNS (which includes massive and intermediate mass
single stars and binaries) and a model that describes the formation and evolution of galaxies. Our group in Brussels is
among the first ones who studied in a consistent way the effects of binaries on GCE. An extended review was published
by De Donder and Vanbeveren (2004). In this paper we first present massive single star yields deduced from
evolutionary calculations performed in Brussels which account for moderate convective core overshooting and for the
effects of recent stellar wind mass loss rate formalisms. For the intermediate mass stars we use synthetic yields which
consider all the dredge up phases during the AGB-phase (note that especially for the estimation of the carbon
enrichment of a galaxy a correct implementation of the latter yields is essential). Secondly, we spent quite some time
to tabulate binary yields as function of primary mass, binary mass ratio and period, using our extended library of
binary tracks and we explain how to implement binaries in a GCE model. To decide whether or not an intermediate mass
binary produces a SN Ia, we separately applied the single-degenerate (SD) model of Hashisu et al. (1999, and references
therein) and the double-degenerate (DD) model. Next, since stellar evolution depends on the initial chemical composition
of the gas out of which stars form, we linked our PNS code to a GCE model. 

Obviously the inclusion of binaries in a GCE model increases significantly the number of parameters. We
made a large number of simulations and when conclusions are formulated, we only restrained those that are
largely independent from the parameter uncertainties. Summarizing:

\begin{itemize}
\item low and intermediate mass binaries enrich less (up to a factor 5) in carbon than single stars do.
This  is due to the fact that the dredge-up phases during the TP-AGB are suppressed by Roche lobe overflow
in most binary systems. 
\medskip
\item of course intermediate mass close binaries are essential in order to understand the SN Ia rate, thus
to  understand the variation of the galactic iron abundance.
\medskip
\item the inclusion of massive close binaries alters the galactic variation of $\alpha$-elements by at most a
factor  2-3. We leave it to the chemical evolutionary community to decide whether or not this difference is
important enough in order to implement binaries in the codes. Notice however that such an implementation is
NOT something simple like changing the yields in existing codes which do not account for the detailed
evolution of a population of binaries. 
\medskip
\item our models predict the appearance of merging degenerate binaries (black hole + neutron star or double 
neutron star binaries) a few million years after the formation of the galactic halo, early enough in order
to explain the observed galactic variation of r-process elements. This conclusion depends essentially on the
physics of case BB evolution after the spiral-in phase of OB+neutron star binaries.
\medskip
\item a chemical evolutionary model of the Galaxy must reproduce the G-dwarf distribution in the disk (the 
number of G-dwarfs as function of iron-abundance). The predicted distribution obviously depends on the
variation of the iron-abundance and therefore depends on whether or not binaries are included, and on
whether the SD and/or DD scenario is adopted in order to calculate the SN Ia rate. Our prediction fits the
observations only when both the SD and DD model produces an SN Ia.

\end{itemize}

Note that Langer et al. (1998) investigated the galactic $^{26}$Al and concluded that massive close binaries may be the
dominant source for this radionuclide.

\noindent
{\bf The rates of different types of SN.} De Donder and Vanbeveren (2004) also investigated the effects of binaries on
the temporal evolution of the galactic SN rates. SN Ia  and most of the SN Ib result from binaries and therefore
binaries are essential in order to calculate SN rates. Summarizing:

\begin{itemize}

\medskip
\item In order to reproduce the present day SN Ia rate of our Galaxy, the progenitor intermediate mass
interacting binary frequency must have been very high, at least 50\%.
\medskip
\item the number of SN II relative to SN Ibc depends significantly on the properties of the massive
binary  population (binary frequency and mass ratio distribution). However, the ratio hardly depends on the
metallicity and therefore the ratio in one particular galaxy varies very slowly as a function of time.
\medskip
\item while the observed number ratio SN Ia/SN Ibc is nearly equal in early and late type spirals
($\sim$1.6), the  observed SN II/SN Ibc ratio is significantly different (about a factor of 2) in both type
of spirals. Since most of the SN Ibc progenitors are binary components, these observed number ratios may
suggest that the massive binary population relative to the intermediate mass binary population is similar in
early and late type spirals but that the overall binary frequency (or the overall binary population) in both
types of spiral galaxies is significantly different.
\medskip
\item the observed high SN Ia/SN Ibc ratio of $\sim$1.6 is difficult to reproduce with either the SD scenario
or the DD scenario separately. The observed ratio is much better approached if both SN Ia scenarios act
together and produce SN Ia. Notice that this conclusion was also reached when the observed and
theoretically predicted G-dwarf distributions are compared (previous subsection).
\medskip
\item for all galaxies together in the observed sample of Cappellaro et al. (1999) the ratio SN II/SN Ibc 
is $\sim$5. Since in the sample all the main morphological types of galaxies are included, this value could
be considered as some cosmological average. It follows from our simulations that to recover this average, we
need a massive binary fraction (on the zero age main sequence) between 40\% and 70\% which could imply that
the cosmological massive binary formation frequency may be of the order of 50\%.
\end{itemize}

\section{3. The population of point-like X-ray sources}

{\bf The standard high mass X-ray binaries.} The scenario for the formation and evolution of the standard high mass
X-ray binaries (HMXBs) proposed by Van den Heuvel and Heise (1972) has been confirmed frequently by detailed binary
evolutionary calculations. We distinguish three X-ray phases: 

\begin{itemize}
\item the OB star is well inside its critical Roche lobe and loses mass by stellar wind. The X-rays are formed when the
compact star accretes mass from the wind (wind fed systems).  
\item The OB star is at the beginning of its RLOF phase and mass transfer towards the compact star starts gently (RLOF
fed systems).
\item The optical star is a Be star and X-rays are emitted when the compact star orbits inside the disk of
the Be star (disk fed systems). 
\end{itemize} 

In the massive binary evolutionary simulations performed with the Brussels code, we detected a
possible fourth phase: when the OB+compact companion binary survives the RLOF-spiral-in-common envelope
phase and the optical star is at the end of its RLOF, burning helium in its core, it transfers mass at a
very moderate rate similar as the rate at the beginning of RLOF. The star is overluminous with respect to
its mass, the surface layers are nitrogen rich and have a reduced surface hydrogen abundance (X $\le$ 0.4).
Possible candidates with an overluminous optical companion are Cen X-3 and SMC X-1. The question here is how
a binary can survive the spiral-in phase? Obviously, some binaries have to survive because we observe double
neutron stars. Theoretically the survival probability becomes larger if one accounts in detail for the
combined action of stellar winds and spiral-in. To illustrate, when after the formation of the compact star,
the binary period is large enough so that an LBV-type or RSG-type stellar wind mass loss can start before
the onset of the spiral-in, the importance of the latter process can be reduced significantly. Our
simulations (with the RSG or LBV wind rates discussed by Vanbeveren et al., 1998a, see also DV) allow to
conclude that it cannot be excluded that some HMXBs are RLOF-fed systems where the optical star is a core
helium burning star at the end of the RLOF.  

Most of the supernova type Ib/c happen in binaries and all HMXBs with a neutron star companion are expected to have
experienced (and survived) such a supernova. Since a WC star is expected to be a type Ic progenitor, evolutionary
calculations predict that the SN shell may contain lots of carbon and oxygen. When this WC star was a binary component
and when the SN shell hits the OB companion star, quite some C and O may be accreted by the latter and abundance
anomalies may be expected. Performing a detailed analysis of the CO abundances in the optical star of Cyg X-1 may be
interesting. An observed overabundance may be an indication that the black hole progenitor experienced a supernova
explosion, may be even a hypernova.

\bigskip

\noindent {\bf Ultra luminous X-ray sources (ULXs).} ULXs are point-like X-ray sources with X-ray luminosities in excess
of 10$^{39}$ ergs$^{-1}$ (Fabbiano, 1986). 

Young supernova remnants (YSNRs) lose rotational energy and part of this is radiated as X-rays. Van Bever and Vanbeveren
(2000) investigated the effects of YSNRs on the X-ray emission of starbursts and concluded that they could dominate the
X-radiation (and identified as a ULX) if the initial rotation period of the neutron star at birth is very small,
smaller than 0.01 sec. Whether or not such small initial periods are real is still a matter of debate.  

Most of the ULXs can be explained by accretion of mass on a 10-20 M$_{\odot}$ stellar mass black hole (BH), but to
explain the most massive ones in the same way, BH masses of 100 to several 100 M$_{\odot}$ are required (Colbert and
Miller, 2004). The latter objects are generally referred to as intermediate mass BHs (IMBHs).   

The BH masses predicted by stellar evolutionary computations depend on the effect of stellar wind mass loss on massive
star evolution, and more specifically on the mass loss during the CHeB-WR-phase. Before 1998, most of the massive star
evolutionary calculations used a WR-mass loss rate formalism which was based on theoretical interpretation of WR
spectra with atmosphere models that assume homogeneity of the stellar wind (Hamann, 1994). However, already in 1996, at
the meeting {\it Wolf-Rayet Stars in the Framework of Stellar Evolution} (eds. J.M. Vreux,
A. Detal, D. fraipont-Caro, E. Gosset, G. Rauw, Universite de Liege) Tony Moffat and John Hillier presented evidence
that WR winds are inhomogeneous implying that the real WR mass loss rates were smaller by at least a factor 2-3. In
1998, we were among the first to perform and publish evolutionary computations of massive stars with such reduced WR
mass loss rates (Vanbeveren et al., 1998a, b). At that time, the evolutionary-referees were not always in favor, to
express it mildly. After, 1998, observational evidence was growing that indeed WR winds are inhomogeneous and that the
rates are lower. Since 2000, everybody is using reduced WR-rates in their evolutionary code. A major consequence of
lower WR mass loss rates is of course the final stellar mass before core collapse. In our 1998 calculations, stars with
a metallicty Z = 0.02 and with an initial mass $\le$ 120 M$_\odot$ end their life with a mass $\le$ 20 M$_\odot$. When
the WR stellar wind mass loss rate is metallicty dependent (as predicted by the radiation driven wind theory, Pauldrach
et al., 1994 among many others), the pre-core collapse mass may be as large as 40-50 M$_\odot$ in small Z environments
(like the SMC for example). 

What about stars with initial mass larger than 120 M$_\odot$? First, let us recall that the
observational evidence for the existence of stars with initial mass $>$ 120 M$_\odot$ is very poor (a
candidate may be the Pistol Star, Figer et al., 1998). However, an important question is the following: if
someway or another a star forms with a mass $>$ 120 M$_\odot$, how does it evolve? Hydrodynamic simulations
(within its limitations) let us suspect that stars with a luminosity close enough to the Eddington value,
will lose mass at very high rate. Luminous blue variables (LBVs) are stars with $\gamma$ = L/L$_{edd}$ close to 1 (Aerts
et al., 2004). One of the most famous LBVs is $\eta$ Car, a star with $\gamma$ $>$ 0.7 and an observed average mass loss
rate $\sim$ 10$^{-3}$ M$_\odot$/yr (high + low state). Evolutionary calculations reveal that already on the
zero age main sequence, a 150 M$_\odot$ has a $\gamma$ = 0.9, a 200 M$_\odot$ even has $\gamma$ = 0.96. This let us
suspect that stars with an initial mass $>$ 120 M$_\odot$ will suffer from a very high stellar wind mass
loss rate already at zero age. The consequences are obvious: when a star is formed with a mass $>$ 120
M$_\odot$, it will very soon evolve into a state where it is almost undistinguishable from a star whose mass
was $\sim$ 120 M$_\odot$ on the zero age main sequence. One may therefore be inclined to conclude that the maximum
stellar BH masses quoted above may be real maxima. 

The foregoing LBV-stellar wind argument and its effect on stellar evolution affect in a critical way the
outcome of N-body dynamical computations of young dense stellar systems, and in particular the formation of
IMBHs by runaway collision (Portegies Zwart et al., 2004). In order to investigate the
LBV-effect on the formation of IMBHs, we recently decided to implement stellar dynamics in our PNS and PSS codes
(Belkus et al., 2005). We use a standard N-body integration technique in order to follow the motion of N objects, an
object can either be a single star or a binary, the interaction of two objects is treated with the chain regularization
method as explained by Mikkola and Aarseth (1993, and references therein) and everything is combined with the PNS and
PSS of starbursts discussed in Van Bever and Vanbeveren (2000, 2003). As a first order simulation, a cluster is
generated with 3000 massive single stars (mass between 10 and 120 M$_{\odot}$) and a King (1966) distribution with
parameters so that the simulation may be appropriate for MGG-11, one of the brightest star clusters in the central
region of M82 (notice that the cluster probably contains a ULX, see Portegies Zwart et al., 2004). When due to real
collisions a star is formed with a mass
$>$ 120 M$_{\odot}$, it evolves with a stellar wind mass loss rate = 10$^{-3}$ M$_{\odot}$/yr. When the mass drops
below 120 M$_{\odot}$, we switch back to normal stellar evolution as it is implemented in our PNS/PSS code. Collision
products are mixed instantaneously and since we follow the pre-collision stars in detail, we calculate the resulting
chemical abundances of the mixed star from first principles. The further evolution of this merger is calculated with
our stellar evolutionary code with the appropriate abundances.  Figure 1 shows a typical simulation. A runaway
collision starts after a few 10$^{5}$ yrs. However, after the major part of the runaway process there is enough time
left for the merger to lose sufficient mass so that it becomes a {\it normal} 120 M$_{\odot}$ star. Notice that the
term {\it normal} may be misleading. Due to the fact that we mix merger products, mergers (and the resulting stars
after thermal relaxation) have nitrogen rich surface layers and their convective core may be significantly larger than
the one of a real normal star (thus also the resulting BH mass at the end of its evolution; to illustrate, the BH mass
of the merger star shown in figure 1 equals 40 M$_{\odot}$, which is a factor 2 larger than the BH mass of a normal 120
M$_{\odot}$ star).     

The main conclusion of our calculations is that in order to study the possibility to form IMBHs in young 
dense stellar systems, a good knowledge of the LBV-type instability in very massive stars and the
resulting mass loss rate is essential. In any case, due to the action of an LBV type instability in stars with a mass
$>$ 120 M$_\odot$, the formation of an IMBH by runaway collision of normal hydrogen burning stars in young dense stellar
systems becomes much more unlikely. 

\bigskip

\begin{figure}[h]

\epsfig{file=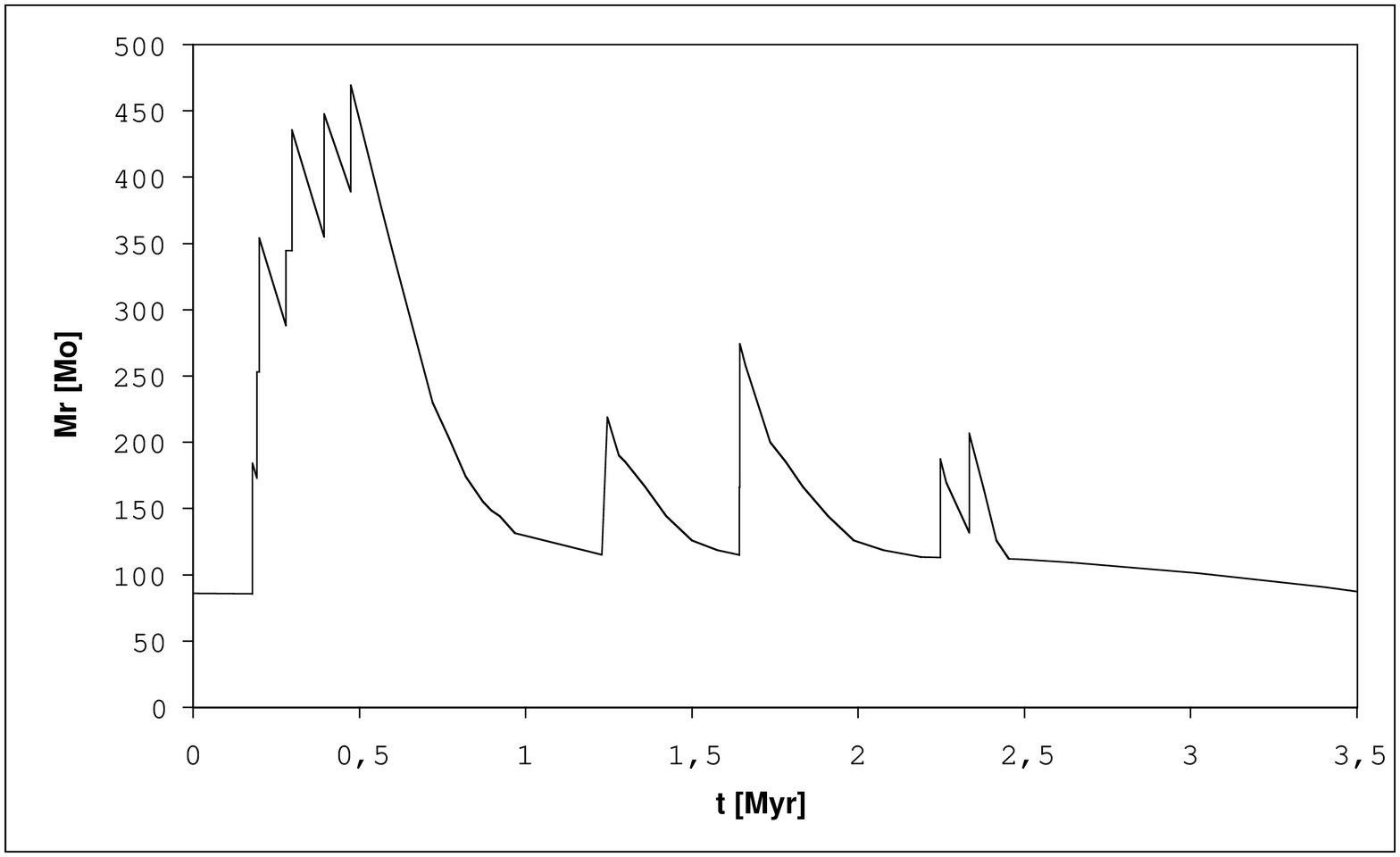, height=7cm,width=12cm}
\caption{The variation of the mass Mr of the most massive star in the cluster. }
\label{fig:fig2}

\end{figure}

\section{4. A UFO-scenario for WR+OB binaries}

The formation of WR+OB binaries in young dense stellar systems may be quite different from the conventional 
binary evolutionary scenario as it was proposed by Van den Heuvel and Heise (1972). Mass segregation in
dense clusters happens on a timescale of a few million years which is comparable to the evolutionary
timescale of a massive star. Within the lifetime of a massive star, close encounters may therefore happen
very frequently. When we observe a WR+OB binary in a dense cluster of stars, its progenitor evolution may be
very hard to predict. Our cluster simulations including the effects of dynamics discussed in the previous section
predict the following unconventionally forming object-scenario ({\it UFO-scenario}) of WR+OB binaries. After 4 million
years the first WR stars are formed, either single or binary. Due to mass segregation, this happens most likely when
the star is in the starburst core. Dynamical interaction with another object becomes probable, especially
when the other object is a binary. In our simulations, we encountered a situation where the WR star (a
single WC-type with a mass = 10 M$_\odot$) encounters a 16  M$_\odot$ + 14 M$_\odot$ circularized binary
with a period P = 6 days. The result of the encounter is the following: the two binary components merge and
the 30 M$_\odot$ merger (which is nitrogen enhanced) forms a binary with the WC star with a period of $\sim$80 days and
an eccentricity e = 0.3. This binary resembles very well the WR+OB binary $\gamma$$^2$Velorum but it is clear that
conventional binary evolution has not played any role in its formation.

\section{5. The effect of rotation on population synthesis}

PNS and PSS depend on stellar evolution and we need to know how evolution is affected by rotation. Rotation 
implies rotational mixing in stellar interiors and it can enhance the stellar wind mass loss
compared to non-rotating stars. This enhancement may be important for stars that are close to the Eddington
limit (LBVs and very massive stars) and therefore rotation may affect indirectly their evolution.

The observed distribution of rotational velocities has been investigated by Vanbeveren et al.  (1998b) and
we illustrated that the majority of the early B-type stars and of the O-type stars are slow rotators, slow
enough to conclude that rotational mixing only plays a moderate role during their evolution (the effect is
similar to the effect of moderate convective core overshooting). In the latter paper we argued that due to
the process of synchronization in binaries, accounting for the observed binary period distribution, a
majority of primaries in massive interacting binaries is expected to rotate slow enough so that the effect
of rotation on their overall evolution is moderate as well.  

The distribution has an extended tail towards very large rotational velocities, i.e. the distribution is
highly asymmetrical which means that in order to study the effect of stellar rotation on population
synthesis (the WR and the O type star population for example), it is NOT correct to use a set of
evolutionary tracks calculated with an average rotational velocity corresponding to the observed average.
This tail obviously demonstrates that there are stars which are rapid rotators. Binary mass
gainers, binary mergers and stellar collision products in young dense stellar environments are expected to
be rapid rotators and thus are expected to belong to the tail. The question however is
whether or not one can approximate their evolution with rotating single star models.  

Due to the dynamo effect, rotation generates magnetic fields (Spruit, 2002) which means that the
evolutionary effect of rotation cannot be studied separately from the effects of magnetic fields. This was
done only since recently (Maeder and Meynet, 2004; see also Norbert Langer in the present proceedings) and
(as could be expected) several of the stellar properties (size of the core, main sequence lifetime, tracks
in the HR diagram, surface abundances etc. ) are closer to those of models without rotation than with
rotation only. Maeder and Meynet notice that single star evolution with rotation only
explains the surface chemistry of the observed massive supergiant population, whereas single star
evolution with rotation and magnetic fields does not. They use this argument to conclude that magnetic fields must be
unimportant. However this argumentation is based on the assumption that most of the massive stars evolve as single
stars do. I can think of at least three other processes which can make stars with non-solar CNO abundances: the  RLOF
process in interacting binaries where the surface layers of the mass loser but also of the mass gainer may becomes
N-enriched, the merger of two binary components due to a highly non-conservative RLOF (common envelope
phase) and last not least, the collision and merger process due to N-body dynamics in young dense stellar
systems. Therefore before an argumentation as the one above has any meaning, one has to consider all these
processes.

All in all, it is my personal opinion that, for the overall synthesis of pre-SN stellar populations, the effect
of rotation is moderate.

\end{document}